\documentclass[12pt]{article}
\usepackage{amssymb,hyperref,graphicx,pbox}

\newcommand{\be}{\begin{equation}}
\newcommand{\ee}{\end{equation}}
\newcommand{\bea}{\begin{eqnarray}}
\newcommand{\eea}{\end{eqnarray}}
\newcommand{\beas}{\begin{eqnarray*}}
\newcommand{\eeas}{\end{eqnarray*}}

\begin{document}
\begin{titlepage}

\medskip

\begin{center}

{\Large Superluminal Propagation on a Moving Braneworld}

\vspace{12mm}

\renewcommand\thefootnote{\mbox{$\fnsymbol{footnote}$}}
Brian Greene${}^{1}$\footnote{brian.greene@columbia.edu},
Daniel Kabat${}^{2,3}$\footnote{daniel.kabat@lehman.cuny.edu},
Janna Levin${}^{4}$\footnote{janna@astro.columbia.edu},
Arjun S.\ Menon${}^{5}$\footnote{armenon@students.chappaquaschools.org}

\vspace{6mm}

${}^1${\small \sl Departments of Physics and Mathematics, Columbia University} \\
{\small \sl 538 West 120th Street, New York, NY 10027, USA}

\vspace{3mm}

${}^2${\small \sl Department of Physics and Astronomy} \\
{\small \sl Lehman College, City University of New York} \\
{\small \sl 250 Bedford Park Blvd.\ W, Bronx, NY 10468, USA}

\vspace{3mm}

${}^3${\small \sl Graduate School and University Center, City University of New York} \\
{\small \sl  365 Fifth Avenue, New York, NY 10016, USA}

\vspace{3mm}

${}^4${\small \sl Department of Physics and Astronomy} \\
{\small \sl Barnard College of Columbia University} \\
{\small \sl New York, NY 10027, USA}

\vspace{3mm}

${}^5${\small \sl Horace Greeley High School} \\
{\small \sl Chappaqua, NY 10514, USA}

\end{center}

\vspace{12mm}

\noindent
We consider a braneworld scenario in the simplest setting, $M_4 \times S^1$, with a 4D Minkowski metric induced on the brane, and establish the possibility of superluminal propagation.
If the brane is at rest, the 4D Lorentz symmetry of the brane is exact, but if the brane is in motion, it is broken globally by the compactification.  By measuring bulk fields, an observer
on the brane sees a slice through a higher-dimensional field profile, which carries an imprint of the extra dimensions even when the brane is at rest.  If the brane is in motion we find that bulk fields can propagate outside the brane
lightcone by a parametrically large amount set by the brane velocity.  We mention observational tests and possible applications to cosmology.

\end{titlepage}

\setcounter{footnote}{0}
\renewcommand\thefootnote{\mbox{\arabic{footnote}}}

\hrule
\tableofcontents
\bigskip
\hrule

\addtolength{\parskip}{8pt}
\section{Introduction\label{sect:intro}}
Imagine living on a brane with a single extra transverse dimension compactified on a circle.  Suppose a high-priority signal needs to be sent by massless messenger between two points on the brane.  Since the bulk spacetime is multiply-connected there is a choice: is it better to send the signal along the brane or launch it into the bulk?  If the brane is at rest, it's clearly optimal to send the signal along the brane.  One might guess that sending along the brane is also optimal if the brane is in motion, but we'll see that this is not the case: signals sent into the bulk can beat signals sent along the brane, by an amount that depends on the brane velocity and can be arbitrarily large.

To describe this in more detail consider a braneworld scenario with a single circular extra dimension.  
We begin with a higher-dimensional Minkowski space with $d$ spacetime dimensions
and split the coordinates as $x^\mu = (t,{\bf x},z)$.  We compactify the $z$ coordinate on a circle of radius $R$ so that
\be
\label{Minkowski}
ds_{\rm bulk}^2 = -dt^2 + \vert d{\bf x} \vert^2 + dz^2 \qquad {\bf x} \in {\mathbb R}^{d-2},\quad z \approx z + 2 \pi R
\ee
This defines a preferred frame in which the identification is purely spatial.  In this frame there is an exact Lorentz symmetry acting on the $(t,{\bf x})$ coordinates, the usual
lower-dimensional Lorentz symmetry that is preserved by Kaluza-Klein compactification.

We will be interested in the effects of brane motion, so instead of working in the preferred frame we consider a frame moving in the compact direction.  We introduce boosted coordinates $(t',{\bf x}',z')$ by setting ${\bf x}' = {\bf x}$ and
\be
\label{boost}
\left(\begin{array}{c} t' \\ z' \end{array}\right) = \left(\begin{array}{cc} \gamma & - \gamma \beta \\ - \gamma \beta & \gamma \end{array}\right) \left(\begin{array}{c} t \\ z \end{array}\right) \qquad\quad
\left(\begin{array}{c} t \\ z \end{array}\right) = \left(\begin{array}{cc} \gamma & \gamma \beta \\ \gamma \beta & \gamma \end{array}\right) \left(\begin{array}{c} t' \\ z' \end{array}\right) \qquad
\ee
Locally these coordinates have the same metric as (\ref{Minkowski}), $ds^2 = -(dt')^2 + \vert d{\bf x} \vert^2 + (dz')^2$, although the identification is no longer purely spatial.  Instead it's given
by transforming $(0,0,2 \pi R)$ into the boosted frame, which means the identification is
\be
\label{identify}
\left(\begin{array}{c} t' \\ {\bf x}' \\ z' \end{array}\right) \approx \left(\begin{array}{c} t' \\ {\bf x}' \\ z' \end{array}\right) + \left(\begin{array}{c} - \gamma \beta 2 \pi R \\ 0 \\ \gamma 2 \pi R \end{array}\right)
\ee
For a braneworld located at $z' = 0$, meaning a moving brane located at position $z = \beta t$ in the preferred frame, the induced metric is
\be
\label{brane}
ds_{\rm brane}^2 = -(dt')^2 + \vert d{\bf x}' \vert^2
\ee
The brane metric is invariant under Lorentz transformations of the $(t',{\bf x}')$ coordinates, and brane-localized matter would presumably respect this symmetry.  This could lead a brane observer
to believe that worldvolume Lorentz invariance is fundamental.  And yet, not only is it not fundamental, but for $\beta \not= 0$ it's not even a symmetry since it doesn't preserve
the identification (\ref{identify}).  The would-be Lorentz symmetry on a moving brane is broken by global effects.

Our goal is to explore a few of the consequences of this global breaking.
For reviews of Lorentz violation see \cite{Liberati:2013xla,Mariz:2022oib} and for related previous work see \cite{Greene:2011fm}.  We'll
examine how bulk signals -- excitations which propagate causally in the bulk -- are perceived by a brane observer.  We'll see that violation of the would-be Lorentz symmetry
on the brane allows for some curious effects.

We start in section \ref{sect:causality} by examining causality, the crudest aspect of signal propagation.  Brane-localized observers could be misled into thinking that causality with respect to the brane metric
is fundamental.  This can be phrased as the requirement that in time $t'$ a signal can spread on the brane according to
\be
\hbox{\rm brane causality:} \qquad \vert {\bf x}' \vert \leq t'
\ee
In truth, only causality with respect to the bulk metric is fundamental.  We'll see that bulk causality allows signals to spread on the brane at a faster rate given by
\be
\label{BulkBound}
\hbox{\rm bulk causality:} \qquad \vert {\bf x}' \vert \leq \gamma t'
\ee
The bulk bound is generically saturated at late times (large $t'$).  In short, causal signals in the bulk become tachyons on the brane: they travel faster than light with respect to the brane metric,
by an amount which can be parametrically large.

At a more refined level, bulk fields propagate in a space with additional compact dimensions.  In section \ref{sect:early}, we discuss the imprint this has on observations
made on the brane.  By measuring a bulk field, a brane observer can directly see a slice through a field propagating in higher dimensions.  The extra dimensions leave an
observable imprint, even if the brane is at rest, simply because the retarded Green's function is dimension-dependent and sensitive to the compactification geometry.  We illustrate this
in a simple example and also show how the signal observed on the brane is modified when the brane is in motion.  This lets us illustrate the full range of imprints of a compact dimension,
from early times to late times.
In section \ref{sect:further}, we mention some observational tests and directions for further development.  Relevant facts about Green's functions are collected in appendix \ref{appendix:Greens}.

\section{Bulk causality on a moving brane\label{sect:causality}}
In this section, we examine bulk causality from the brane point of view and show that it allows for apparent faster-than-light travel on the brane.
We first show this from simple geometric considerations involving light cones then study the effect in more detail in terms of retarded Green's functions.

\subsection{Light cones\label{sect:geometry}}
The geometric argument runs as follows.  Imagine a source of light at the origin $t = {\bf x} = z = 0$.  In the covering space
where the $z$ coordinate is unwrapped this source corresponds to an infinite series of image charges at
\be
t_w = {\bf x}_w = 0 \qquad z_w = 2 \pi R w \qquad w \in {\mathbb Z}
\ee
At time $t$, the light cones of these image charges form a series of circles.
\be
\vert {\bf x} \vert^2 + (z - z_w)^2 = t^2
\ee
In the boosted frame, the image charges are located at
\be
t_w' = - \gamma \beta 2 \pi R w \qquad {\bf x}_w' = 0 \qquad z_w' = \gamma 2 \pi R w
\ee
and their light cones expand as
\be
\vert {\bf x}' \vert^2 + (z' - z_w')^2 = (t' - t_w')^2
\ee
Noting that $t_w' = - \beta z_w'$ this can also be written as
\be
\label{LightCones1}
\vert {\bf x}' \vert^2 + (z' - z_w')^2 = (t' + \beta z_w')^2
\ee

\begin{figure}
\begin{center}
\includegraphics[height=5.5cm]{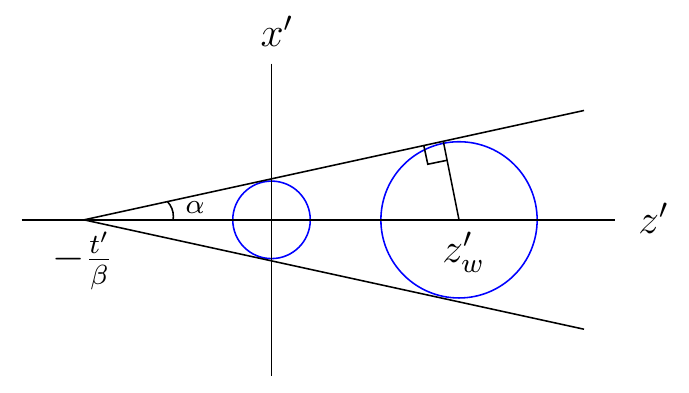}
\caption{Blue circles: light cones produced by image charges on a slice of constant $t'$.  At time $t'$, the light cone centered at $z_w'$ has radius $t' + \beta z_w'$.  The envelope of the light cones forms a cone along the $z'$
axis with tip at $z' = - t' / \beta$ and opening angle $\alpha = \sin^{-1} \beta$.\label{fig:cone}}
\end{center}
\end{figure}

\begin{figure}
\begin{center}
\includegraphics[height=7.5cm]{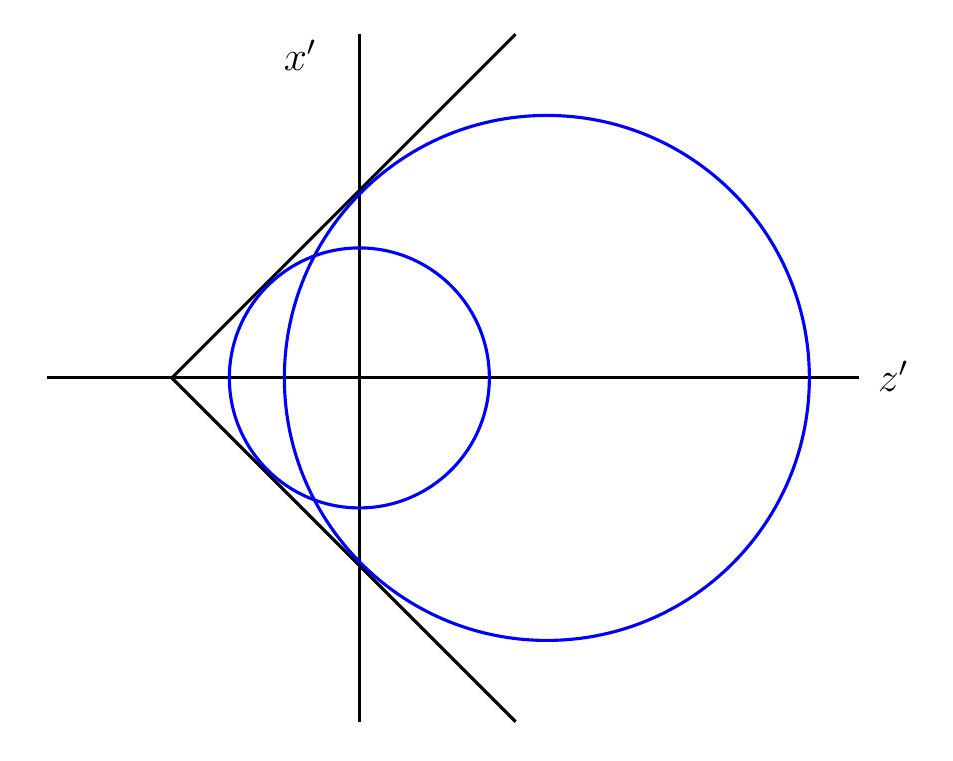}
\caption{Blue circles: light cones produced by image charges on a slice of constant $t'$.  The brane is at $z' = 0$.  The wavefront centered to the right has spread further along the brane than the one centered at the origin.\label{fig:cone2}}
\end{center}
\end{figure}

As shown in Fig.\ \ref{fig:cone}, the envelope forms a cone along the $z'$ axis with the tip of the cone at $z' = - {1 \over \beta} t'$ and an opening angle $\alpha$ satisfying
\be
\sin \alpha = {\hbox{\footnotesize radius of light cone} \over \hbox{\footnotesize tip-to-center distance}} = \beta
\ee
At any given $z'$, the envelope has an extent in the directions parallel to the brane
\be
\vert {\bf x}' \vert = (z' + {1 \over \beta} t') \tan \alpha = {t' + \beta z' \over \sqrt{1 - \beta^2}}
\ee
In particular on a brane located at $z' = 0$ the envelope expands according to
\be
\label{violation}
\vert {\bf x}' \vert = \gamma t'
\ee
This exceeds the speed limit suggested by the induced brane metric by a factor of $\gamma$.  The reason for the superluminal propagation can be seen in Fig.\ \ref{fig:cone2}.  Wavefronts produced by image
charges to the right hit the brane and eventually spread further along the brane than the wavefront produced by the image charge at the origin.  The effect becomes more pronounced as $\beta$ increases, since the opening angle
$\alpha \rightarrow \pi/2$ as $\beta \rightarrow 1$.

For later reference, it's useful to note the time at which the various image charges first becomes visible to observers on the brane.  The $w^{th}$ image charge
can first be seen on the brane at position ${\bf x}' = z' = 0$ at a time
\be
\label{Doppler}
t' = t_w' + \vert z_w' \vert = \left\lbrace\begin{array}{cc}
2 \pi R w \sqrt{1 - \beta \over 1 + \beta} & \hbox{\rm for $w > 0$} \\[5pt]
- 2 \pi R w \sqrt{1 + \beta \over 1 - \beta} & \hbox{\rm for $w < 0$}
\end{array}\right.
\ee
These are nothing but the usual relativistic Doppler formulas.  Image charges in the direction of motion are seen at a blue-shifted frequency while image charges behind the observer are red-shifted.
It's worth noting that as $\beta \rightarrow 1$ all of the image charges with $w > 0$ appear instantly on the brane: the time delay one might expect, due to the travel time around the compact dimension, gets
completely Doppler-shifted away.

\begin{figure}
\centerline{\includegraphics[width=10cm]{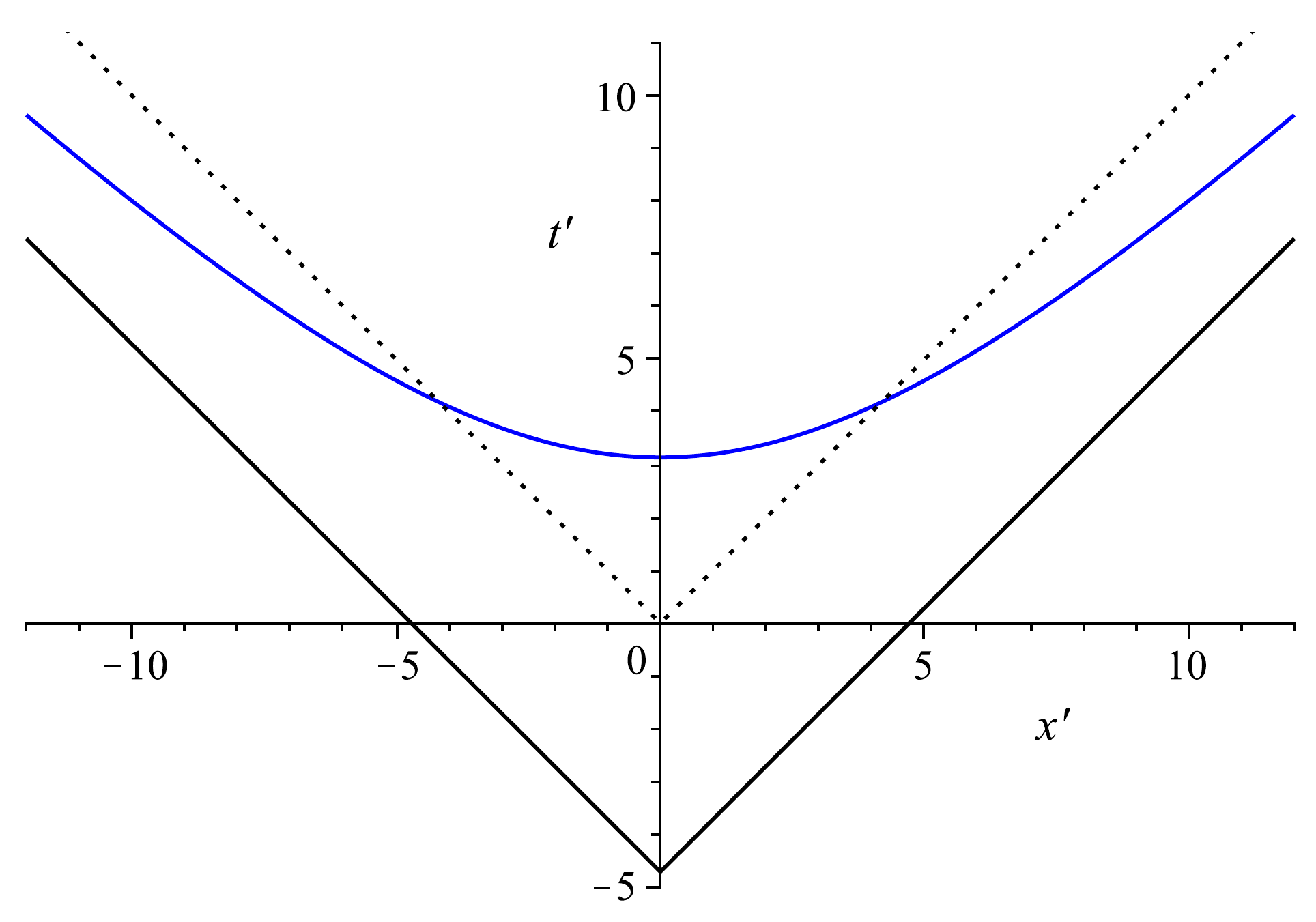}}
\caption{An image charge with $w > 0$ produces the spacelike hyperbola shown in blue.  The signal first appears on the brane at a time given by the Doppler formula (\ref{Doppler}) and asymptotically approaches
the lightcone with origin at $t' = - \beta z_w'$ shown in solid black.  It eventually spreads outside the brane lightcone shown in dotted black.  Plot is for $w = 1$, $R = 1$, $\beta = 0.6$.\label{fig:hyperbola}}
\end{figure}

We can also be more explicit about how the light cones of the image charges appear on the brane.  Setting $z' = 0$ in (\ref{LightCones1}) we see that the light cone of the $w^{th}$ image charge spreads along the brane
according to
\be
\label{LightCones2}
\vert {\bf x}' \vert^2 = (t' + \beta z_w')^2 - (z_w')^2
\ee
This defines a spacelike hyperboloid on the brane, illustrated in Fig.\ \ref{fig:hyperbola}, with the special case $w = 0$ being the brane lightcone $\vert {\bf x}' \vert = t'$.  The hyperboloid is asymptotic to a light
cone originating from $t' = - \beta z_w'$, so all image charges produce signals that asymptotically
expand at the speed of light on the brane.  If the brane velocity is non-zero, the image charges with $w > 0$ have origins at $t' < 0$ and eventually spread outside the brane light-cone, while image
charges with $w < 0$ produce hyperboloids that forever remain inside the brane light-cone.

The expression (\ref{LightCones2}) is useful for understanding how a particular image charge appears on the brane, but to understand the envelope effect discussed above, it's better to re-write (\ref{LightCones2}) in the equivalent form
\be
\label{LightCones3}
\vert {\bf x}' \vert^2 = (\gamma t')^2 - (\gamma \beta t' - 2 \pi R w)^2
\ee
In this form, we see that for any given $t'$ the leading-edge signal on the brane is produced by the image charge with $w \approx \gamma \beta t' / 2 \pi R$.  For this particular image charge, the final term vanishes
and the signal has spread a distance $\vert {\bf x}' \vert \approx \gamma t'$, showing that a succession of image charges with increasing values of $w$ are responsible for the envelope discussed above.

\subsection{Green's functions\label{sect:Greens}}
The effect can be understood in more detail from a study of the retarded Green's function for a bulk field.
In parallel with our split of the coordinates as $x^\mu = (t,{\bf x},z)$ we split the momenta as $k^\mu = (\omega,{\bf k},q)$.

Consider a bulk scalar field of mass $m$.  We denote the retarded Green's function $G^{(d)}_R(t,{\bf x},z)$ where $d$ is the number of spacetime dimensions and $R$ is the radius of the circle.  We start from the retarded Green's function in the
covering space which satisfies
\beas
&& \left(-\partial_\mu\partial^\mu + m^2\right) G^{(d)}_\infty(t,{\bf x},z) = \delta(t) \delta^{d-2}({\bf x}) \delta(z) \\
&& G^{(d)}_\infty(t,{\bf x},z) = 0 \quad \hbox{\rm for $t < 0$}
\eeas
This Green's function has a representation
\be
\label{Gd}
G^{(d)}_\infty(t,{\bf x},z) = \int {d\omega \over 2 \pi} {d^{d-2}k \over (2\pi)^{d-2}} {dq \over 2\pi} \, {e^{-i \omega t} \, e^{i {\bf k} \cdot {\bf x}} \, e^{i q z} \over -\omega^2 + \vert {\bf k} \vert^2 + q^2 + m^2}
\ee
where the $\omega$ contour is deformed to pass above the poles.  Although it's not obvious from the integral representation, causality requires
\be
G^{(d)}_\infty(t,{\bf x},z) = 0 \quad \hbox{\rm for $t < \sqrt{\vert {\bf x} \vert^2 + z^2}$}
\ee

To compactify the $z$ direction we introduce an image sum which can also be thought of as a sum over winding numbers.
\be
\label{image}
G^{(d)}_R(t,{\bf x},z) = \sum_{w \in {\mathbb Z}} G^{(d)}_\infty(t,{\bf x},z - 2 \pi R w)
\ee
Alternatively, we can compactify the $z$ direction by making the sum over momentum modes discrete.\footnote{To see the equivalence of the winding and momentum forms consider
$\sum_{w \in {\mathbb Z}} e^{i q (z - 2 \pi R w)}$.  This is periodic in $z$ so can be expanded in a Fourier series $\sum_n c_n e^{i n z / R}$ with coefficients given by
\[
c_n = {1 \over 2 \pi R} \int_0^{2 \pi R} dz e^{-i n z / R} \sum_{w \in {\mathbb Z}} e^{i q (z - 2 \pi R w)} = {1 \over 2 \pi R} \int_{-\infty}^\infty dz e^{-i n z / R} e^{i q z} = {1 \over R} \delta\big(q - {n \over R}\big)
\]
This leads to the identity
\[
\sum_{w \in {\mathbb Z}} e^{i q (z - 2 \pi R w)} = {1 \over R} \sum_{n \in {\mathbb Z}} \delta\big(q - {n \over R}\big) e^{i n z / R}
\]
Using this for the sum in (\ref{image}) leads to (\ref{KK}).}
\be
\label{KK}
G^{(d)}_R(t,{\bf x},z) = {1 \over 2 \pi R} \sum_{n \in {\mathbb Z}} \int {d\omega \over 2 \pi} {d^{d-2}k \over (2\pi)^{d-2}} \, {e^{-i \omega t} \, e^{i {\bf k} \cdot {\bf x}} \, e^{i n z / R} \over -\omega^2 + \vert {\bf k} \vert^2 + \big({n \over R}\big)^2 + m^2}
\ee

Now we can see how the field responds to a source at the origin.  At early times $t < 2 \pi R$, only the $w = 0$ term in the winding sum contributes so
\be
G^{(d)}_R(t,{\bf x},z) = G^{(d)}_\infty(t,{\bf x},z) \qquad \hbox{\rm for $t < 2 \pi R$}
\ee
This is an exact statement, enforced by causality.  It means that at early times, the field propagates as though the theory had full $d$-dimensional Lorentz invariance.

As time goes by, more and more image charges will contribute.  At late times, the winding sum becomes continuous, $\sum\limits_{w \in {\mathbb Z}} \rightarrow \int dw$.  Making this replacement in (\ref{image}) and using (\ref{Gd}) we see that the winding sum leads to
\be
\int dw \, e^{-i q 2 \pi R w} = {1 \over R} \delta(q)
\ee
This freezes the $q$ integral and leads to the late-time behavior
\be
G^{(d)}_R(t,{\bf x},z) \rightarrow {1 \over 2 \pi R} G^{(d-1)}_\infty(t,{\bf x})
\ee
An equivalent statement is that the late-time behavior is dominated by the Kaluza-Klein mode with $n = 0$.
At late times the $z$ dependence drops out and the behavior is governed by the Green's function in $d-1$ dimensions.

The late-time Green's function has the expected $(d-1)$-dimensional Lorentz symmetry of the preferred
frame.  In particular at late times signals propagate causally in the preferred frame with
\be
\vert {\bf x} \vert < t
\ee
But clocks on a moving brane run slow.  The coordinates appropriate to a brane observer are obtained by setting $t = \gamma t'$ and ${\bf x} = {\bf x}'$, and in these coordinates the bound becomes
\be
\vert {\bf x}' \vert < \gamma t' 
\ee
To a brane observer it appears that bulk fields can propagate superluminally, in agreement with (\ref{violation}).

\section{Bulk fields on a moving brane\label{sect:early}}
We've seen that the retarded Green's function undergoes crossover from the Green's function in $d$ non-compact dimensions at early times to the Green's function in $d-1$ non-compact dimensions at late times.  The timescale for the crossover is set by $2 \pi R$ in the preferred frame, the moment at which bulk fields first notice the compactification.  Here we study the crossover in more detail and
point out some of the observational consequences.

Brane motion is not essential to most of the discussion in this section.  The crossover phenomenon is present even for a brane at rest, so we begin by illustrating it in that context, then
present the straightforward generalization to a moving brane.

Closely related phenomena have been studied in the literature.  For a brane at rest the crossover in the static (purely spatial) Green's function is responsible for the famous
modification of the Newtonian potential at short distances in the large extra dimension scenario \cite{Arkani-Hamed:1998jmv}, while modifications to the Newtonian potential on a moving brane
have been studied in \cite{Greene:2011fm}.  In a sense our goal here is merely to extend the analysis from a static Green's function to a retarded Green's function.  This will allow us to make contact with our previous discussion of
causality on a moving brane.

Let's imagine that an observer on the brane has access to a source $J(x)$ that can excite a bulk field $\phi(x)$ (and for definiteness, we set $d = 5$). The source should be well-localized, both on the brane and in the
compact dimension. The bulk field will then propagate in all dimensions according to the retarded Green's function.
\be
\label{field1}
\phi(x) = \int d^5x' \, G^{(5)}_R(x-x') J(x')
\ee
We have in mind the simplest setting where the the bulk field is massless and the compact dimension is a circle of radius $R$.  The appropriate Green's function $G^{(5)}_R$
is given by the image sum (\ref{image}),
\be
\label{G5R}
G^{(5)}_R(t,{\bf x},z) = \sum_{w \in {\mathbb Z}} G^{(5)}_\infty(t,{\bf x},z - 2 \pi R w)
\ee
where the Green's function in the covering space $G^{(5)}_\infty$ is given in (\ref{GdMassless3}).
\be
\label{G5}
G^{(5)}_\infty(t,{\bf x},z) = {i \over 8 \pi^2} \theta(t) \Bigg[{1 \over \left(\vert {\bf x} \vert^2 + z^2 - (t - i \epsilon)^2\right)^{3 / 2}} - {1 \over \big(\vert {\bf x} \vert^2 + z^2 - (t + i \epsilon)^2\big)^{3 / 2}}\Bigg] \hspace{2cm}
\ee
Here $\epsilon \rightarrow 0^+$ serves to define the singularities in the Green's function.  At late times $t \gg 2 \pi R$ we expect to have
\be
G^{(5)}_R(t,{\bf x},z) \rightarrow {1 \over 2 \pi R} G^{(4)}_\infty(t,{\bf x}) = {1 \over 4 \pi^2 R} \, \theta(t) \, \delta(t^2 - \vert {\bf x} \vert^2)
\ee
by the arguments of section \ref{sect:Greens}.  This should be understood as convergence in the sense of a distribution.

To present explicit results, we need to choose a source function $J(x)$.  This introduces a great deal of freedom.  A convenient choice is simply to keep $\epsilon$
small but non-zero in (\ref{G5}).  This smooths out the Green's function and defines a corresponding source $J_\epsilon(x)$ through
\be
(\partial_t^2 - \nabla_{\bf x}^2 - \partial_z^2) \, G^{(5)}_{\infty}(t,{\bf x},z) = J_\epsilon(x)
\ee
As $\epsilon \rightarrow 0^+$, we're guaranteed that $J_\epsilon(x) \rightarrow \delta^d(x)$, but for finite $\epsilon$ the source is smeared over a length scale $\sim \epsilon$.

\begin{figure}[t]
\centerline{\hbox{\hspace{-4mm}\includegraphics[height=9cm]{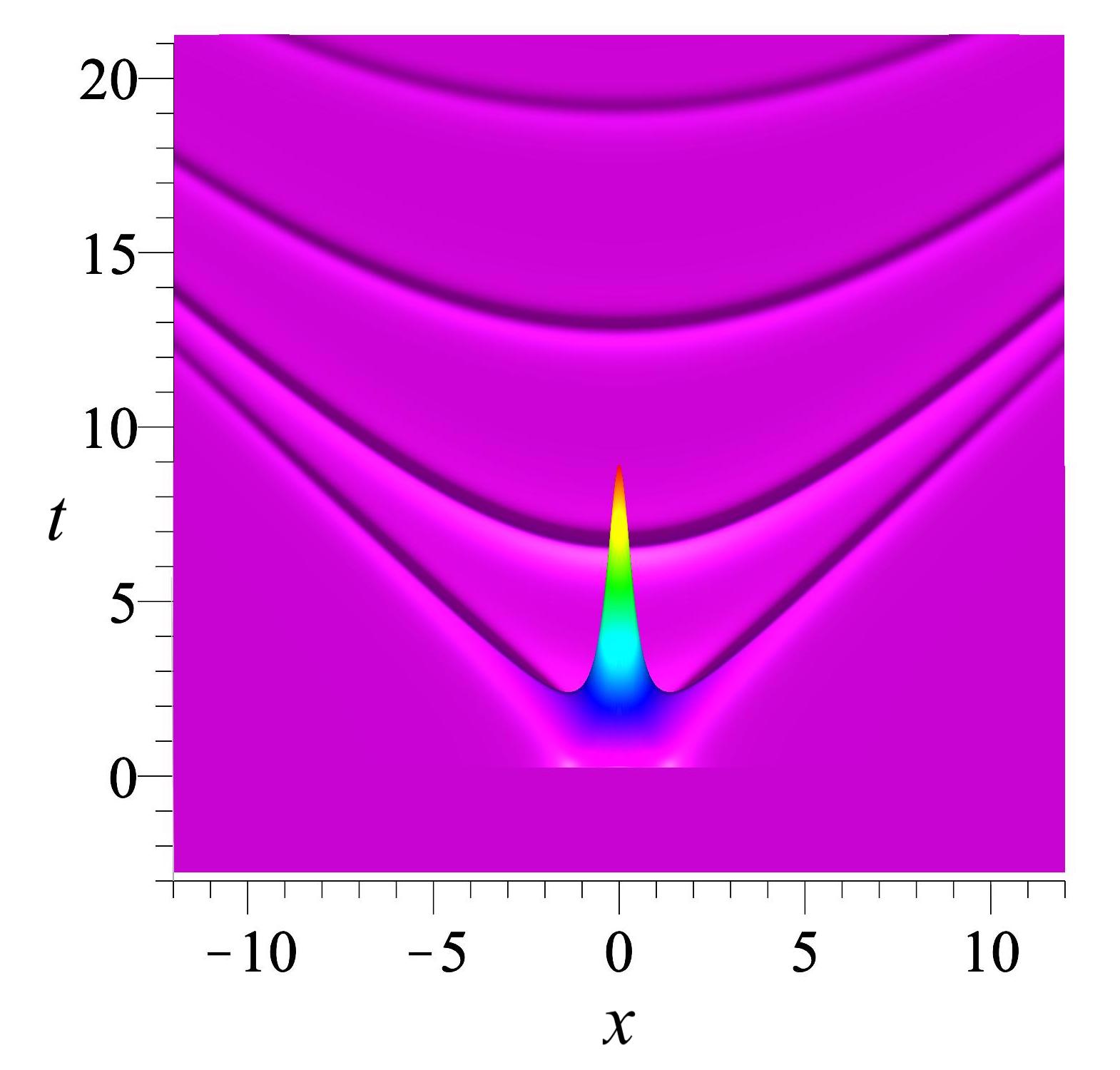} \hspace{2mm} \includegraphics[height=9cm]{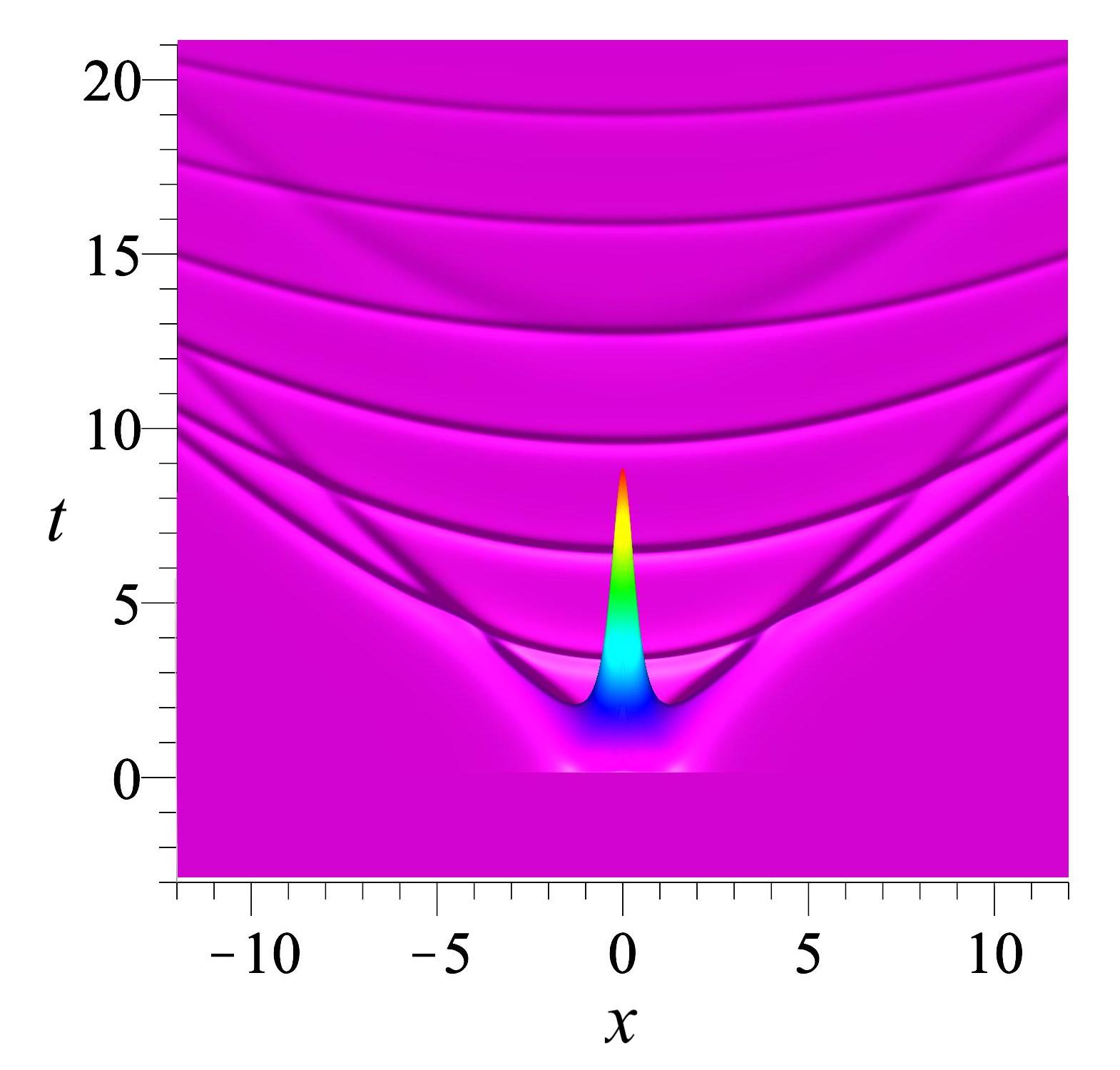}}}
\caption{The field profile on a brane at rest (left panel) and a brane in motion with $\beta = 0.6$ (right panel).  Note the resemblance to Fig.\ \ref{fig:hyperbola}.
When the brane is in motion the image charges in the direction of motion are blue-shifted
and spread outside the lightcone of the induced metric on the brane.  In both figures the compactification radius is $R = 1$ and the source is smeared with $\epsilon = 0.5$.\label{fig:profile}}
\end{figure}

At this point, we present some numerical results.  We imagine that the source $J_\epsilon(x)$ is centered at the origin, $t = {\bf x} = z = 0$, and we start by considering a stationary brane
located at $z = 0$.  An observer on the brane could measure the field profile produced by the source.  This can be obtained by setting $z = 0$ in (\ref{G5R}).
\be
\phi_{\rm stationary}(t,{\bf x}) = {i \over 8 \pi^2} \theta(t) \sum_{w \in {\mathbb Z}} {1 \over \big(\vert {\bf x} \vert^2 + (2 \pi R w)^2 - (t - i \epsilon)^2\big)^{3/2}} + {\rm c.c.}
\ee
This is illustrated in the left panel of Fig.\ \ref{fig:profile}.
A few comments are in order.  First, since the brane is stationary, there is an unbroken Lorentz symmetry that acts on the $(t,{\bf x})$ coordinates.  The field would look the same to
any inertial observer on the brane and the naive brane causality bound $\vert {\bf x} \vert \leq t$ is obeyed.  However at early times an observer on the brane sees a slice through the
5D Green's function which -- unlike the 4D Green's function -- is non-vanishing inside the future light cone (the so-called Hadamard tail, see \cite{Aruquipa:2022rwr} for a recent discussion).
Thus even at early times, a brane observer can see an imprint of the
extra dimension.  At $t = 2 \pi R$, the observer can start to see image charges, an even more dramatic signal of the extra dimension.  At late times, the image charges accumulate
and the signal approaches the 4D Green's function.   In other words, the signal approaches what one would expect from Kaluza-Klein reduction, which means that for observers on a stationary brane
the imprint of the extra dimension goes away at late times.

The situation becomes more interesting if the brane is in motion.  Suppose the brane is moving in the $z$ direction, $z = \beta t$.  In the boosted frame (\ref{boost}), the moving
brane is located at $z' = 0$.  Coordinates on the brane $(t',{\bf x}',z'=0)$ correspond to coordinates in the bulk via the inverse Lorentz transformation
\be
t = \gamma t' \qquad {\bf x} = {\bf x}' \qquad z = \gamma \beta t'
\ee
To get the field profile on a moving brane, we simply plug these coordinates into our general expression for the Green's function (\ref{G5R}).  This gives
\be
\label{PhiMoving}
\phi_{\rm moving}(t',{\bf x}') = {i \over 8 \pi^2} \theta(t') \sum_{w \in {\mathbb Z}} {1 \over \big(\vert {\bf x}' \vert^2 + (\gamma \beta t' - 2 \pi R w)^2 - (\gamma t' - i \epsilon)^2\big)^{3/2}} + {\rm c.c.}
\ee
which is illustrated in the right panel of Fig.\ \ref{fig:profile}.

Again, a few comments are in order.  The term with $w = 0$ has 5D Lorentz invariance and isn't sensitive to the motion of the brane, so at early times we have
\be
\hbox{\rm early times:} \quad \phi_{\rm moving}(t',{\bf x}') = {i \over 8 \pi^2} \theta(t') {1 \over \big(\vert {\bf x}' \vert^2 - (t' - i \epsilon)^2\big)^{3/2}} + {\rm c.c.}
\ee
At early times, the would-be 4D Lorentz symmetry on the brane is respected and the naive brane causality bound $\vert {\bf x}' \vert \leq t'$ is obeyed.
The field is still a slice through a 5D Green's function, hence non-zero in the future light cone, however there is no sign that the brane is in motion.
But after a time $t' = 2 \pi R \sqrt{1 - \beta \over 1 + \beta}$ the image charges start to become visible, and these do violate the would-be Lorentz
symmetry of the brane.  In particular, at late times, when the winding number becomes continuous, there is always an image charge with $\gamma \beta t' - 2 \pi R w \approx 0$.
From (\ref{PhiMoving}) we can see that these image charges have light cones that spread on the brane at a rate that saturates the bulk causality bound $\vert {\bf x}' \vert \leq \gamma t'$.

\section{Further developments\label{sect:further}}
In this paper, we've examined the behavior of bulk fields from the perspective of a brane observer.  For a brane at rest, worldvolume Lorentz transformations are an exact symmetry, but even so a brane observer can see an
imprint of the extra dimensions since the bulk field profile is sensitive to the compactification.  If the brane is in motion the worldvolume Lorentz symmetry is broken by global effects, with the curious consequence that bulk fields can propagate faster than the speed of light that is induced on the brane.  These phenomena are most likely to be relevant if there are large extra dimensions \cite{Antoniadis:1990ew,Arkani-Hamed:1998jmv,Antoniadis:1998ig}, a scenario recently revisited in \cite{Montero:2022prj}.

Related studies have been carried out in the
literature.  In particular \cite{Greene:2011fm} considered the Kaluza-Klein tower seen by a moving brane, obtaining the dispersion relation from the brane point of view
\be
\omega' = \gamma \sqrt{\vert {\bf k}' \vert^2 + m^2 + \left({n \over R}\right)^2} - {\gamma \beta n \over R}
\ee
The effects of brane motion are encoded in this dispersion relation.   Note that worldvolume Lorentz invariance is violated when the brane is in motion.  Also a Kaluza-Klein mode propagates on the brane with group velocity
\be
v_g = {d \omega' \over d k'} = {\gamma \vert {\bf k}' \vert \over \sqrt{\vert {\bf k}' \vert^2 + m^2 + \left({n \over R}\right)^2}}
\ee
Not surprisingly, at large $\vert {\bf k}' \vert$, the group velocity saturates the propagation speed we found for brane-localized sources at late times,  $v_g \rightarrow \gamma$ as $\vert {\bf k}' \vert \rightarrow \infty$.

We conclude with a few further developments which can also be viewed as directions for future work.

\noindent
{\em Observational tests} \\
To detect the effects we have discussed, a brane observer must be able to interact with a bulk field.  Various candidates for bulk fields have been proposed including gravity \cite{Arkani-Hamed:1998jmv,Randall:1999ee}
and sterile neutrinos \cite{Dienes:1998sb,Arkani-Hamed:1998wuz, Montero:2022prj}.  The Hadamard tail of the higher-dimensional Green's function \cite{Aruquipa:2022rwr} would provide a clear signal of extra dimensions,
as would the so-called fireworks associated with image charges \cite{Arkani-Hamed:1998jmv} which are modified if the brane is in motion \cite{Greene:2011fm}.

More generally, on a moving brane, the worldvolume Lorentz symmetry is broken by global effects.  Lorentz violation has been investigated extensively, for reviews see \cite{Liberati:2013xla,Mariz:2022oib}.
As a direct and dramatic signal of Lorentz violation, we focus on the possibility of faster-than-light travel on the brane.  Here the constraints from multi-messenger astrophysics \cite{Addazi:2021xuf,QG-MM}
can be very stringent.  Assuming that light travels on the brane with speed $c$ while a bulk field propagates with speed $c + \Delta c$, the velocity bound (\ref{BulkBound}) directly translates to
\be
{\Delta c \over c} = \gamma - 1
\ee
The speed of electron antineutrinos compared to photons was measured in the time-of-flight experiment conveniently provided by supernova SN1987a which led to bounds $\Delta c / c \lesssim 10^{-8}$
\cite{Stodolsky:1987vd} and $\Delta c / c \lesssim 2 \times 10^{-9}$ \cite{Longo:1987gc}.  These bounds were later improved to $\Delta c / c \lesssim 10^{-10}$ \cite{Ellis:2008fc,Liberati:2013xla}.  The speed of gravitational
waves compared to photons was likewise measured using the binary neutron star merger GW170817 / GRB 170817A with the result \cite{Abbott_2017}
\be
-3 \times 10^{-15} \leq {\Delta c \over c} \leq +7 \times 10^{-16}
\ee
If we entertain the possibility that these are bulk degrees of freedom, the bounds on brane velocity become very strict.

\noindent
{\em Applications to cosmology} \\
Although observational tests suggest stringent constraints on brane motion today, the situation could be different in the early universe.  This raises an interesting possibility for addressing the
horizon problem.

To illustrate the idea, we consider the same static bulk geometry with an extra dimension compactified on a circle of radius $R$, but we allow the brane velocity to depend on time, $\beta \rightarrow \beta(t)$.
Assuming the brane velocity changes adiabatically the induced metric on the brane is approximately Minkowski, and for brane-localized matter the particle horizon -- the distance a particle can travel from
time $t_1'$ to time $t_2'$ -- is simply
\be
d_H^{\rm \, brane} = t_2' - t_1'
\ee
However bulk fields travel at speed $\gamma$, so for bulk fields the particle horizon is
\be
d_H^{\rm \, bulk} = \int_{t_1'}^{t_2'} \gamma(t'') \, dt''
\ee
If the brane is in motion we have $d_H^{\rm \, bulk} > d_H^{\rm \, brane}$.  In this way bulk fields could in principle thermalize regions which a brane observer might think are out of causal contact.  The mechanism has the flavor of variable-speed-of-light cosmology, with different effective propagation speeds for different species of particles, although in our
case the various speeds have a unified higher-dimensional description.  For a recent discussion of variable-speed-of-light cosmology see \cite{Bassett:2000wj}.  As a direction for further work it would be interesting
to extend the above analysis beyond a static bulk geometry and consider cosmological metrics on the brane and in the bulk.  In particular it would be interesting to see if it can be applied to the ekpyrotic scenario \cite{Khoury:2001wf}.  

\noindent
{\em Other compactifications} \\
We considered the simplest possibility of a single extra dimension compactified on a circle, but it would be interesting to consider the effects of brane motion in more general and realistic
compactifications.  Simple toroidal compactifications should be straightforward to analyze as they amount to replacing the
winding sum in (\ref{image}) with a sum over a lattice.  More ambitiously
it would be interesting to consider the effects of brane motion in realistic string or F-theory compactifications, perhaps including the scenario outlined in \cite{Montero:2022prj}.
An intermediate step might be to consider brane motion on orbifolds or K3.

\bigskip
\goodbreak
\centerline{\bf Acknowledgements}
\noindent
BG is supported by DOE award number DE-SC0011941.
DK is supported by U.S.\ National Science Foundation grant PHY-2112548. JL is supported in part by the Tow Foundation.

\noindent
ASM contributed to the work reported in section \ref{sect:geometry}.

\appendix
\section{Retarded Green's functions\label{appendix:Greens}}
Green's functions for the wave equation may be a venerable topic \cite{Courant-Hilbert}, but properties of Green's functions are not so familiar especially in position space
in higher dimensions.  For this reason we collect a few results in this appendix.  A pedagogical review for physicists has been prepared by Balakrishnan
\cite{Balakrishnan1,Balakrishnan2} and lecture notes for mathematicians are available from Oh \cite{sjoh}.

We begin with the Fourier representation of the retarded Green's function.
\be
\label{GdMassless}
G^{(d)}_\infty(t,{\bf x}) = \int {d\omega \over 2 \pi} {d^{d-1}k \over (2\pi)^{d-1}} \, {e^{-i \omega t} \, e^{i {\bf k} \cdot {\bf x}} \over -\omega^2 + \vert {\bf k} \vert^2}
\ee
This is the Green's function in $d$ non-compact dimensions.  (We've lumped all the spatial coordinates into ${\bf x}$.)  To simplify the expressions that follow, we've
set the mass to zero.  To produce a retarded Green's function the contour for the $\omega$ integral is deformed to pass above the poles at $\omega = \pm \vert {\bf k} \vert$.
This ensures that the Green's function satisfies
\bea
\label{DiffEqn}
&& (\partial_t^2 - \nabla_{\bf x}^2) \, G^{(d)}_\infty(t,{\bf x}) = \delta(t) \delta^{d-1}({\bf x}) \\[5pt]
\nonumber
&& \quad G^{(d)}_\infty(t,{\bf x}) = 0 \quad \hbox{\rm for $t < 0$}
\eea

For $t < 0$, the $\omega$ contour can be closed in the upper half plane and the Green's function vanishes.  For $t > 0$ it can be closed in the lower half plane where it
encircles the poles and leads to
\be
\label{GdMassless2}
G^{(d)}_\infty(t,{\bf x}) = i \int {d^{d-1}k \over (2\pi)^{d-1}} \left({e^{-i \vert {\bf k} \vert t} \, e^{i {\bf k} \cdot {\bf x}} \over 2 \vert {\bf k} \vert} - {e^{i \vert {\bf k} \vert t} \, e^{i {\bf k} \cdot {\bf x}} \over 2 \vert {\bf k} \vert}\right)
\ee
This expresses the Green's function as a difference of two distributions.  The first term is the boundary value of a function analytic in the lower half of the complex $t$ plane,
which means it can be defined by a $t \rightarrow t - i \epsilon$ prescription.  The second term is the complex conjugate of the first.  It is analytic in the upper half of the complex $t$ plane
and can be defined by $t \rightarrow t + i \epsilon$.

To proceed, it's convenient to set ${\bf x} = 0$ so the momentum integral is spherically symmetric; the dependence on ${\bf x}$ can be restored later by Lorentz invariance.  For the first term in (\ref{GdMassless2}), this gives
\bea
\nonumber
& & i \int {d^{d-1} k \over (2\pi)^{d-1}} \, {e^{-i \vert {\bf k} \vert (t - i \epsilon)} \over 2 \vert {\bf k} \vert} \\[5pt]
& = & {i \, {\rm vol}(S^{d-2}) \over 2 (2 \pi)^{d-1}} \int_0^\infty dk \, k^{d-3} e^{-i k (t - i \epsilon)} \\[5pt]
\nonumber
& = & {\Gamma\big({d-2 \over 2}\big) \over 4 \pi^{d/2}} \, {i \over (i t + \epsilon)^{d-2}}
\eea
In the second line, we used ${\rm vol}(S^{d-2}) = {2 \pi^{d-1 \over 2} / \Gamma\big({d-1 \over 2}\big)}$ and in the final line we made use of some $\Gamma$-function identities.
Restoring Lorentz invariance and recalling that the Green's function is only non-zero for $t > 0$, this means
\be
\label{GdMassless3}
G^{(d)}_\infty(t,{\bf x}) = {i \Gamma\big({d-2 \over 2}\big) \over 4 \pi^{d/2}} \theta(t) \Bigg[{1 \over \big(\vert {\bf x} \vert^2 - (t - i \epsilon)^2\big)^{d-2 \over 2}} - {1 \over \big(\vert {\bf x} \vert^2 - (t + i \epsilon)^2\big)^{d-2 \over 2}}\Bigg]
\ee
A few key features are now transparent.  The Green's function vanishes at spacelike separation, where the $i \epsilon$ prescription isn't needed and the two terms in (\ref{GdMassless3}) exactly cancel.  When the spacetime dimension $d$ is even, the Green's function only has poles,
meaning it only has support on the future light cone, but when $d$ is odd there is a branch cut and it also has support in the interior of the future light cone.

For future reference, it's convenient to work in terms of proper time $\tau = \sqrt{t^2 - \vert {\bf x} \vert^2}$ and write the Green's function as
\be
G^{(d)}_\infty(t,{\bf x}) = \theta(t) \left(\left.H^{(d)}(\tau)\right\vert_{t \rightarrow t - i \epsilon} + {\rm c.c.}\right)
\ee
where
\bea
\label{HdIntegral}
H^{(d)}(\tau) & = & i \int {d^{d-1} k \over (2\pi)^{d-1}} \, {e^{-i \vert {\bf k} \vert \tau} \over 2 \vert {\bf k} \vert} \\
\nonumber
& = & {i \over 4 \pi^{d/2}} \Gamma\Big({d - 2 \over 2}\Big) \big(-\tau^2\big)^{2 - d \over 2}
\eea
There's an amusing relation between Green's functions in different dimensions, since it's straightforward to check that $H^{(d)}(\tau)$ satisfies
\be
\label{recursion1}
H^{(d+2)}(\tau) = - {1 \over 2 \pi (d-1)} {d^2 \over d\tau^2} H^{(d)}(\tau)
\ee
(a second derivative with respect to $\tau$) as well as
\be
\label{recursion2}
H^{(d+2)}(\tau) = {1 \over \pi} {d \over d(\tau^2)} H^{(d)}(\tau)
\ee
(a first derivative with respect to $\tau^2$).

Explicit Green's functions in low dimensions are listed in table \ref{table:Greens}.  For $d = 1,\,2$ the momentum integral in (\ref{GdMassless2}) diverges in the IR.  In these cases,
one can obtain the Green's function by introducing a mass $m$ as a regulator and sending $m \rightarrow 0$ at the end of the calculation; alternatively one can check
directly that the differential equation (\ref{DiffEqn}) is satisfied.  The result in $d = 3$ follows from the discontinuity across the branch cut in (\ref{GdMassless3}).
To obtain the result in $d = 4$ we used the identity ${1 \over x + i \epsilon} = {\rm PV} \, {1 \over x} - i \pi \delta(x)$.

\begin{table}
\begin{center}
\begin{tabular}{c|l}
\raisebox{3mm}{\pbox{10cm}{\hspace*{0.5mm} spacetime \\ dimension $d$}} \quad & \quad \raisebox{1.5mm}{$G^{(d)}_\infty(t,{\bf x})$} \\
\hline
1 & \quad $t \, \theta(t) \phantom{\Bigg]}$ \\[10pt]
2 & \quad ${1 \over 2} \, \theta(t) \, \theta(t^2 - x^2) = {1 \over 2} \, \theta(t + x) \, \theta(t - x)$ \\[10pt]
3 & \quad ${1 \over 2 \pi \sqrt{t^2 - \vert {\bf x} \vert^2}} \, \theta(t) \, \theta(t^2 - \vert {\bf x} \vert^2)$ \\[10pt]
4 & \quad ${1 \over 2 \pi} \, \theta(t) \, \delta(t^2 - \vert {\bf x} \vert^2)$
\end{tabular}
\end{center}
\caption{Retarded Green's functions for a massless field in $d$ non-compact dimensions.\label{table:Greens}}
\end{table}

However, our main interest is in higher dimensions, where the Green's function (\ref{GdMassless3}) becomes increasingly singular on the light cone.  To make sense of the singularity,
note that our real goal is to find the field produced by a source $J(x)$.  That is, we must regard the Green's function as a distribution and interpret the integral
\be
\label{field2}
\phi(x) = \int d^dx' \, G^{(d)}_\infty(x-x') J(x')
\ee
To do this, we use the recursion relation (\ref{recursion2}).  The aim is to write $G^{(d)}$ as a differential operator acting on a lower-dimensional hence less singular Green's function
and integrate by parts.  To do this, we must promote ${d \over d(\tau^2)}$ to a vector field on spacetime.  There is no unique way to do this, and the most obvious choice (making the
vector field orthogonal to hypersurfaces of constant $\tau$) is badly behaved.\footnote{Constant-$\tau$ hypersurfaces are asymptotically null so the unit normal vector asymptotically has
divergent components.}
Instead it's convenient to note that for any function of $\tau^2 = (t-t')^2 - \vert {\bf x} - {\bf x'} \vert^2$ we have
\be
{\partial \over \partial t'} f(\tau^2) = - 2 (t - t') {d \over d(\tau^2)} f(\tau^2)
\ee
This lets us write the recursion relation (\ref{recursion2}) in the form
\be
H^{(d+2)}(\tau) = - {1 \over 2\pi (t - t')} {\partial \over \partial t'} H^{(d)}(\tau)
\ee
Although not manifestly Lorentz covariant, this can be used in (\ref{field2}) to obtain for example
\be
\phi(x) = {1 \over 2 \pi} \int d^dx' \, G^{(d-2)}_\infty(\tau) \left({1 \over t - t'} {\partial \over \partial t'} J(x') + {1 \over (t-t')^2} J(x')\right)
\ee

We conclude with a few additional results from the literature.  Up to an overall normalization the massless Green's function is fixed by Lorentz and scale invariance and can be written in
any dimension as a fractional derivative of a $\delta$-function \cite{sjoh}.
\be
G^{(d)}_\infty(\tau) = \theta(t) {1 \over 2 \pi^{d-2 \over 2}} \left({d \over d(\tau^2)}\right)^{d - 4 \over 2} \delta(\tau^2)
\ee
The extension to massive fields is straightforward but a bit cumbersome.  For example, in place of (\ref{HdIntegral}) we have ($\omega_k = \sqrt{\vert {\bf k} \vert^2 + m^2}$)
\be
H^{(d)}(\tau) = i \int {d^{d-1} k \over (2\pi)^{d-1}} \, {e^{-i \omega_k \tau} \over 2 \omega_k} = - {1 \over 4} \left(m \over 2 \pi \tau\right)^{d-2 \over 2} H^{(1)}_{d-2 \over 2}(-m\tau)
\ee
where $H^{(1)}_{d-2 \over 2}$ is a Hankel function, and the recursion relation (\ref{recursion1}) is replaced with \cite{Balakrishnan1,Balakrishnan2}
\be
H^{(d+2)}(\tau) = - {1 \over 2 \pi (d-1)} \left({d^2 \over d\tau^2} + m^2\right) H^{(d)}(\tau)
\ee

\providecommand{\href}[2]{#2}\begingroup\raggedright\endgroup

\end{document}